\documentclass{ws-procs10x7} 

\def\Journal#1#2#3#4{{#1} {\bf #2}, #3 (#4)}
\def\gsim{\stackrel{>}{_\sim}}

\begin{document}

\title{
Recent Progress Towards a Cost-Effective Neutrino Factory Design}
\author{Daniel M. Kaplan} 

\address{Physics Division, Illinois Institute of Technology, 3101 S. Dearborn Street, Chicago, 
Illinois, 60616 USA\\
E-mail: kaplan@iit.edu\\[0.15in]
{\rm for the Neutrino Factory and Muon Collider Collaboration}
} 

\twocolumn[\maketitle\abstract
{A Neutrino Factory, sending $>10^{20}$ decay neutrinos per year from a high-energy stored muon beam towards remote detectors, has been suggested as the ultimate tool for precision measurement of the neutrino mixing matrix. Following two rounds of design studies that focused primarily on feasibility, the latest such study has begun the process of cost optimization. New ideas include muon `phase rotation' using high-frequency rf cavities and rapid muon acceleration in non-scaling FFAG rings. The world-wide Neutrino Factory R\&D effort is aimed at demonstrating the feasibility of the required techniques by the end of this decade. If this effort is successful and the next round of neutrino experiments confirms the need for a Neutrino Factory, a timely decision to proceed to Neutrino Factory construction will then be possible.}]

\section{Introduction} 
It has been suggested\cite{Geer} that decay neutrinos from a stored high-energy
muon beam (in a `Neutrino Factory') are the ultimate tool\cite{ultimate}  for the study of neutrino oscillations and their possible role in baryogenesis via {\em CP}-violating
neutrino mixing.\cite{baryon} An alternate approach (`Beta Beams') uses decay neutrinos from stored high-energy beams of beta-unstable isotopes.\cite{Zucchelli}
The Neutrino Factory concept 
 arose out of earlier work on a possible muon collider.\cite{Ankenbrandt}  A desirable intensity goal for such a facility is a few\,$\times10^{20}$
neutrinos/year. 

Two Neutrino Factory fea\-si\-bi\-li\-ty stu\-dies\cite{FS1,FS2} were carried out in the US during 1999--2001 (commissioned by Fermilab and Brookhaven National Laboratory, respectively), spearheaded by members of the Neutrino Factory and Muon Collider Collaboration (Muon Collaboration).\cite{MC} (In parallel, less detailed studies were carried out at CERN\cite{CERN-study} and in Japan.\cite{Japan-study}) The US studies each included sufficient engineering effort to produce a realistic cost estimate, so that the key `cost drivers' could be identified for further R\&D. Nevertheless, these studies were focused primarily on demonstrating the {\em feasibility} of a facility that could deliver the desired performance;  {cost} was viewed as a secondary consideration. While the design of Feasibility Study I missed the above intensity goal by about an order of magnitude,\cite{FS1}  the improvements of Feasibility Study II (FS2) increased the intensity per proton-on-target by a factor of 6,\cite{FS2} and an additional factor of 4 can be achieved with a more powerful proton source.\cite{Alsharoa}

In the fall of 2003, the American Physical Society commissioned a year-long {\em Study on the Physics of Neutrinos}.\cite{APS-Study} In that context, the Muon Collaboration re-optimized the design, taking advantage of recent conceptual advances to maintain the FS2 performance, with reduced cost and with the possibility to utilize both muons and antimuons simultaneously.\cite{FS2a}

\section{Neutrino Factory Overview}
Muon production by hadron beams is quite efficient, since charged pions decay essentially 100\% into muons. However, as daughters of secondary products of the initial collision, muons are naturally produced into a large phase space, while affordable existing
acceleration technologies have small transverse acceptances, and consequently favor input beams with small transverse size and angular divergence. This mismatch could in principle be alleviated by
developing new, large-aperture, acceleration techniques\cite{Japan-study} or by `cooling' the muon
beam to reduce its size and divergence; we have in fact opted to do both. Given the 2.2\,$\mu$s muon lifetime, only one cooling technique
is fast enough:  ionization cooling, in which muons repeatedly traverse an energy-absorbing
medium, alternating with accelerating devices, within a strongly focusing magnetic
lattice.\cite{cooling,Neuffer83} 
In this way the transverse-momentum spread of the beam is reduced while the average longitudinal momentum remains constant, thus the beam's transverse degrees of freedom are cooled.

\begin{figure}[t]
\vspace{-.95in}
\centerline{\hspace{.05in}
\includegraphics[bb=0 0 650 800,width=3in]{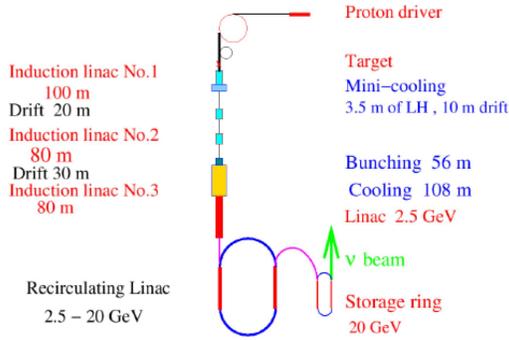}}
\vspace{-.95in}
\caption{Layout of Neutrino Factory from US Feasibility Study II.\protect\cite{FS2}
\label{fig:FSII}
}
\end{figure}

A Neutrino Factory thus consists of a high-power proton source and target facility, an (optional) muon-cooling system, fast accelerators, and finally a storage ring in which the muons decay, with long straight sections aimed at near and remote detectors. A schematic layout of the FS2 design is shown in Fig.~\ref{fig:FSII}. A point to note is the small footprint of the facility, which easily fits on the sites of existing laboratories such as Fermilab, BNL, and CERN.

\section{Physics Reach}

The key advantage of both Beta Beam and Neutrino Factory facilities over conventional neutrino beams is that they provide intense, collimated, and well-understood beams of electron neutrinos and antineutrinos. Neutrino oscillation is then signalled by $\stackrel{(-\!-)}{\nu_\mu}$ charged-current interactions in the far detector, producing muons or antimuons. Backgrounds to the identification of muons in large detectors are at least two orders of magnitude lower than those for electron detection. The three `holy grails' of neutrino-oscillation physics\,---\,measurements of the $\theta_{13}$ mixing angle, the neutrino mass hierarchy, and the {\em CP}-violating phase, $\delta$, of the mixing matrix\,---\,are thus achievable with a Neutrino Factory for $\theta_{13}$ values below $10^{-4}$. No other technique has this sensitivity.

The priority of a Neutrino Factory in the world's future High Energy Physics program depends on measurements to be made in upcoming experiments. If $\theta_{13}$ is sufficiently large ($\gsim$0.05) it will be measured in currently proposed reactor experiments. For $\theta_{13}$ values just below the current experimental upper limit ($\theta_{13}\sim0.1$), Superbeam and Beta Beam experiments taken together may be able to make the three measurements mentioned above. Even then, for the ultimate in precision a Neutrino Factory may eventually be needed. For $\theta_{13}\gsim0.05$, `medium-energy' Beta Beams may approach Neutrino Factory sensitivity to $\delta$ within a factor of a few. However, determination of the neutrino mass hierarchy relies on the differing interactions of neutrinos and antineutrinos with matter as they pass through the earth, and the higher energy and longer baselines of Neutrino Factory experiments confer a substantial advantage. Moreover, by providing (from $\mu^+$ decays) simultaneous beams of electron neutrinos and muon antineutrinos (and {\it vice versa} for $\mu^-$ decays), a Neutrino Factory will offer unrivaled power to sort out systematic effects, degeneracies, and ambiguities due to correlations among the measurements.\cite{FS2a}

\section{`Feasibility Study IIa'}

The APS {\em Study on the Physics of Neutrinos}\cite{APS-Study} presented the opportunity for the FS2 design to be updated. Since there were neither time nor resources for engineering, the `FS2a' design was conceived as a perturbation on that of FS2; thus the costs of the revised components can be estimated with some confidence by scaling the corresponding FS2 costs. The combined performance of the `front end' (everything up to the acceleration section) is the same as that of the FS2 design: $0.170\pm0.006$ muons per proton-on-target, delivered within the acceptance of the acceleration. This represents a doubling of the FS2 performance, since muons of both charges are transmitted, in opposite crests of the radio-frequency (rf) waveform. 

The components of the FS2a design are briefly summarized in the following; details may be found in Ref.~\cite{FS2a}.
\begin{itemize}
\item {\bf Proton Driver:}
As in FS2, the proton source is taken as the 24\,GeV Brookhaven AGS, upgraded to  1--4\,MW of proton-beam power. (As shown in other studies,\cite{FS1,CERN-study,RAL-study} a high-power proton accelerator of a different design could serve equally well.)

\item {\bf Target and Capture:}
A high-power mercury-jet target is immersed in a 20\,T solenoidal field to capture charged pions produced in proton-nucleus interactions. The high magnetic field at the target (produced by a superconducting outer coil with a resistive insert) is smoothly tapered down to 1.75\,T, which is then maintained through the bunching and phase-rotation sections. 

\item {\bf Bunching and Phase Rotation:}
Before cooling, the muons' longitudinal phase space must be matched to the acceptance of the cooling channel. This requires `phase rotation', in which the energy spread of the beam is reduced and its time spread increased. In FS2, phase rotation was done before bunching, using (expensive) induction linacs. The new approach employs rf cavities, at frequencies that decrease along the channel from 333 to 201\,MHz, to first bunch and then phase-rotate the muon beam. Besides reduced cost, this rf phase rotation has the advantage that it simultaneously transports muons of both charges.

\item {\bf Cooling:}
The FS2 cooling channel employed liquid-hydrogen energy absorbers in a superconducting-solenoid focusing lattice and reduced the transverse normalized rms emittance of the muon beam by a factor of about 6, from 12 to 2\,mm$\cdot$rad. In FS2a, solid LiH absorbers are used instead, cooling the transverse emittance from 17\,mm$\cdot$rad to about 7\,mm$\cdot$rad. (This takes place at a central muon momentum of 220 MeV/$c$.) An unusual feature of the normal-conducting rf cavities used  to restore the muons' longitudinal momentum between absorbers is that their apertures are closed by low-$Z$ windows to reduce both the surface electric fields and power costs. The LiH absorbers (sandwiched by thin Be foils) in fact perform double duty, serving as the cavity windows as well. Costs are reduced relative to FS2 by use of a shorter, simpler cooling channel with less focusing strength, and hence fewer, smaller superconducting solenoids.

\item {\bf Acceleration:}
The muon acceleration was the most expensive part of the FS2 design and is substantially reconceived in Study IIa. The FS2 design featured a superconducting linac to bring the muon energy up to 2.5\,GeV, followed by a single, racetrack-shaped, superconducting recirculating linear accelerator (RLA) that accelerated the beam to 20\,GeV. In FS2a, the initial linac accelerates to 1.5\,GeV and is followed by a `dogbone' RLA, accelerating to 5\,GeV. The muons then enter two cascaded non-scaling fixed-field alternating-gradient (FFAG) rings, the first accelerating to 10 and the second to 20\,GeV. This system has twice the transverse acceptance as that of FS2 and so is well matched to the larger emittance of the beam emerging from the FS2a cooling channel. As shown in Fig.~\ref{fig:accel}, transfer lines are incorporated for muons of both charges.

\begin{figure*}[t]
\centerline{\includegraphics[width=3.5in]{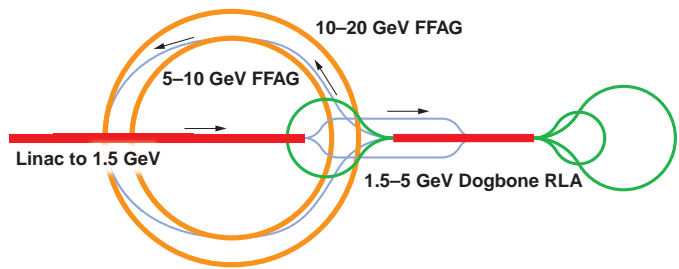}}
\caption{FS2a acceleration layout.
\label{fig:accel}}
\end{figure*}

\item {\bf Storage Ring:}
The FS2 design is used in FS2a as well:  a compact, racetrack-shaped, superconducting storage ring 
in which $\approx 35$\% of the stored muons decay toward a detector located some 3000\,km 
from the ring. Muons survive for roughly 500 turns. The muon and antimuon bunch trains are arrayed within the ring such that the arrival of their decay products at the detector can be distinguished by timing ($\Delta t\sim100\,$ns).
\end{itemize}

\section{Cost Reduction}

The hardware cost of the FS2a design is estimated at 67\% of the FS2 cost. Both of these costs include the Proton Driver and Target facility. If (as is not unlikely) these are built in support of an earlier Neutrino Superbeam experiment, their costs should not be attributed to the Neutrino Factory; excluding these the FS2a hardware cost is 60\% of the FS2 cost, or about \$900M. If the Proton Driver and target are included the cost rises by about \$300M. (These are preliminary, `unloaded' figures and also depend on the accounting approach used, which may differ significantly among the various world regions in which such a facility might be built.)

There is additional scope for cost reduction, possibly at some minor penalty in performance. Further work is needed to optimize the cost and performance; in a global minimization the cost and performance of the detector will also play a role. It would also be desirable to do a `bottom-up' cost estimate for the new FS2a design, and for any further variations. It is likely that this work will become part of a `World Design Study', whose planning is now getting under way.

\section{Crucial Demonstration Experiments}

It is evident from the above that Neutrino Factory feasibility and performance depend on a number of extrapolations beyond the current state of the accelerator art. An international R\&D program is in progress to certify the reliability of these extrapolations.
\begin{itemize}

\item{\bf Targetry:}
A mercury jet with the needed 20\,m/s jet velocity is under construction, to be tested in a normal-conducting 15\,T solenoid in the nTOF beamline at CERN.\cite{targetry} The instantaneous power deposition from the beam (180\,J/g) will match that at a 4\,MW Proton Driver, but, to allow time for the magnet to recover between pulses, the apparatus will be pulsed only occasionally. The experiment is planned for completion within the next few years.

\item{\bf Muon Cooling:}
The first phase of an international Muon Ionization Cooling Experiment (MICE) has recently been approved for construction at the Rutherford Appleton Laboratory.\cite{MICE} First beam is planned for the spring of 2007. In the first phase of MICE, two solenoidal precision tracking spectrometers will be assembled and cross-calibrated in the muon beam. In a subsequent (not yet fully funded) phase, a short section (one lattice cell) of cooling channel based on the FS2 design will be built up and operated between the two spectrometers. The expected 10\% emittance reduction will be measured to 1\% of itself, for a definitive demonstration of the feasibility and performance of ionization-cooling hardware. A variety of absorber materials and lattice optics configurations will be tested, so that performance predictions can be made with confidence not only for the FS2 cooling lattice but for others (e.g., that of FS2a) as well.

\item{\bf Acceleration:}
The proposed `non-scaling' FFAG acceleration is a new approach and entails some unconventional beam dynamics.\cite{FFAG} A scaled-down demonstration of such a non-scaling FFAG using an electron beam has been discussed.\cite{eFFAG} Funding for such an effort remains to be identified, but it is hoped that the project will progress rapidly during the next few years.

\end{itemize}

\section{Summary and Outlook}

A world-wide R\&D program is in progress to establish the feasibility of a stored-muon-beam Neutrino Factory. If all goes as planned, by about 2010 the demonstration experiments and World Design Study will have been completed. All will then be in readiness for a construction decision to be taken, should the coming generation of neutrino experiments indicate (as now seems likely) that a Neutrino Factory is required to answer the key questions at the heart of neutrino mixing.

\section*{Acknowledgments}

This work was supported in part by  the US Dept.\ of Energy, the National Science Foundation, 
and the UK Particle Physics and Astronomy Research Council.

\end{document}